\documentclass[lettersize,journal]{IEEEtran}
\usepackage{amsmath, amsfonts, amssymb}
\usepackage{algorithmic}
\usepackage{algorithm}
\usepackage{array}
\usepackage[caption=false,font=normalsize,labelfont=sf,textfont=sf]{subfig}
\usepackage{textcomp}
\usepackage{stfloats}
\usepackage{url}
\usepackage{verbatim}
\usepackage{graphicx}
\usepackage{cite}
\usepackage[flushleft]{threeparttable}
\usepackage{multirow}
\usepackage{makecell}
\usepackage{bm}
\usepackage{booktabs}
\usepackage{tikz}
\usepackage{float}
\usepackage{hyperref}
\usepackage{arydshln}

\begin{document}

\title{SpeechX: Neural Codec Language Model as a Versatile Speech Transformer}

\author{Xiaofei Wang, Manthan Thakker, Zhuo Chen, Naoyuki Kanda, Sefik Emre Eskimez, Sanyuan Chen, \\ Min Tang, Shujie Liu, Jinyu Li, Takuya Yoshioka
\thanks{X. Wang is the corresponding author (email: xiaofei.wang@microsoft.com).}
\thanks{X. Wang, M. Thakker, Z. Chen, N. Kanda, S. E. Eskimez, M. Tang, J. Li, and T. Yoshioka are with Microsoft Corporation, Redmond, WA 98052, USA. S. Chen and S. Liu are with Microsoft Research Asia, Beijing, China.}}



\maketitle

\begin{abstract}
Recent advancements in generative speech models based on audio-text prompts have enabled remarkable innovations like high-quality zero-shot text-to-speech. However, existing models still face limitations in handling diverse audio-text speech generation tasks involving transforming input speech and processing audio captured in adverse acoustic conditions. This paper introduces SpeechX, a versatile speech generation model capable of zero-shot TTS and various speech transformation tasks, dealing with both clean and noisy signals. SpeechX combines neural codec language modeling with multi-task learning using task-dependent prompting, enabling unified and extensible modeling and providing a consistent way for leveraging textual input in speech enhancement and transformation tasks.
Experimental results show SpeechX's efficacy in various tasks, including zero-shot TTS, noise suppression, target speaker extraction, speech removal, and speech editing with or without background noise, achieving comparable or superior performance to specialized models across tasks.  See \url{https://aka.ms/speechx} for demo samples. 
\end{abstract}

\begin{IEEEkeywords}
Speech generation, audio-text input, multi-task learning, zero-shot text-to-speech, noise suppression, target speaker extraction, speech editing, speech removal
\end{IEEEkeywords}

\section{Introduction}
\label{sec:intro}
The technology of generative models has undergone rapid and transformative advancements in various machine learning applications, encompassing text~\cite{NEURIPS2020_1457c0d6,openai2023gpt4}, vision~\cite{{9878449}}, and audio~\cite{borsos2023audiolm}. These advancements have had significant implications for both the industry and society at large. Notably, generative models using multi-modal input have emerged as a remarkable innovation~\cite{NEURIPS2022_960a172b,wang2022git,ruiz2023dreambooth,yang2023icode,rubenstein2023audiopalm,agostinelli2023musiclm}.

In the speech domain, one prominent speech generation task that leverages audio-text input is zero-shot text-to-speech (TTS). Zero-shot TTS involves converting a given text into speech with the voice characteristics and speaking style of a desired talker by using only a brief audio sample of that person. Early studies in zero-shot TTS employed fixed-dimensional speaker embeddings~\cite{pmlr-v162-casanova22a, NEURIPS2018_6832a7b2, 9054535, casanova21b_interspeech}. This approach limited their usage to TTS alone and did not adequately support speaker cloning capabilities.

In contrast, recent approaches have embraced more generic formulations, such as masked speech prediction~\cite{le2023voicebox} or neural codec language modeling~\cite{wang2023neural,kharitonov2023speak,huang2023makeavoice,zhang2023speak}. These novel approaches directly utilize the target speaker's audio without compressing it into a fixed-dimensional representation. Consequently, these models have not only achieved remarkable zero-shot TTS performance but also demonstrated additional capabilities, including voice conversion~\cite{le2023voicebox, huang2023makeavoice} and speech editing~\cite{le2023voicebox}. This enhanced flexibility holds tremendous promise for unlocking new possibilities in speech generation models.

However, despite their impressive achievements, these recent generative models still have certain limitations, particularly when it comes to addressing various audio-text-based speech generation tasks involving transforming input speech. For instance, existing speech editing models~\cite{bai20223,jiang2023mega} are restricted to handling clean signals only, lacking the ability to modify spoken content while preserving background sounds. Additionally, to perform denoising, the model discussed in~\cite{le2023voicebox} necessitates the noisy signal to be surrounded by clean speech segments, imposing significant constraints on its practical applications. 
In the context of transforming non-clean speech, another particularly useful task is target speaker extraction~\cite{wang2018voicefilter, 8736286, 9746962}. Target speaker extraction involves extracting the voice of a desired speaker from a speech mixture containing multiple talkers. The desired speaker can be specified using a short voice recording of that individual. Despite its potential significance as discussed in~\cite{zmolikova2023neural}, this task remains unaddressed by existing generative speech models.

It is noteworthy that traditional approaches to speech enhancement tasks, such as denoising and target speaker extraction, have relied on regression models for faithful signal recovery. However, these prior methods typically required distinct expert models for each task, which is not ideal, given the potential diversity of acoustic disturbances~\cite{serra2022universal}. Furthermore, there has been a lack of comprehensive audio-text-based speech enhancement models that leverage reference transcriptions to generate intelligible speech, except for limited studies focusing only on particular speech enhancement tasks~\cite{kinoshita15_interspeech, 9053182}.

\begin{figure*}[t!]
  \centering
  \includegraphics[width=\textwidth]{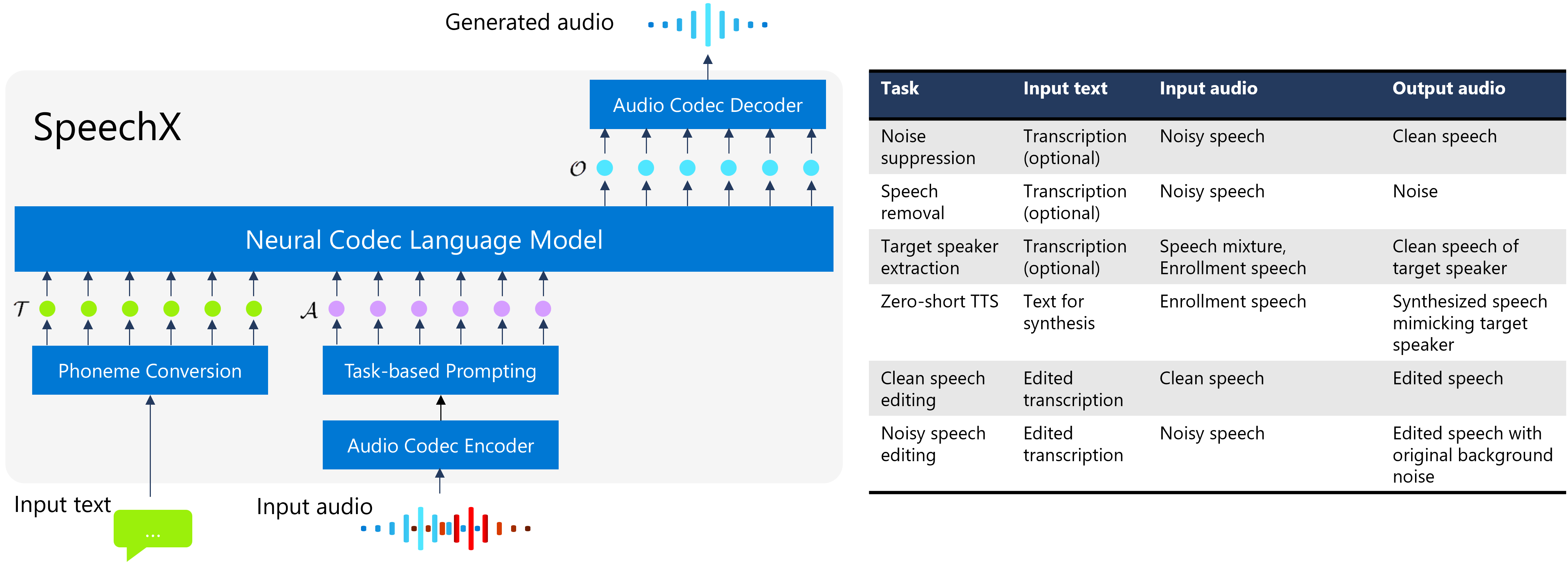}
      \caption{Overview of SpeechX. SpeechX handles multiple audio-text-based speech generation tasks, including noise suppression, speech removal, target speaker extraction, zero-shot TTS, clean speech editing, and noisy speech editing, by using neural codec language model conditioned on the text and acoustic token stream. Text input is optional for some tasks.}
      \label{fig:SpeechX}
\end{figure*}

Given the aforementioned considerations and the successful precedents in other domains, the creation of audio-text-based generative speech models unifying generation and transformation capabilities assumes crucial research importance. These models should possess an overarching capability to tackle a diverse array of speech generation tasks. We propose that such models should be equipped with the following key properties:
\begin{itemize}
\item \textbf{Versatility:} Similar to unified or foundation models developed in other machine learning domains, the unified audio-text-based generative speech models must handle a wide range of tasks involving speech generation from audio and text inputs. These tasks should encompass not only zero-shot TTS but also various forms of speech transformation, including speech enhancement and speech editing, to name a few.

\item \textbf{Robustness:} It is essential for the unified models to exhibit robustness to various acoustic distortions since they are likely to be applied in acoustically challenging environments. By ensuring reliable performance, these models can be deemed highly usable in real-world scenarios where background sounds are prevalent.

\item \textbf{Extensibility:} The unified models must employ flexible architectures, allowing for seamless extensions of task support. One approach to achieving this involves accommodating additional elements, such as input tokens or extra modules. Such flexibility will empower the models to adapt to future speech generation tasks efficiently.
\end{itemize}

In pursuit of this objective, this paper introduces a versatile speech generation model capable of performing multiple tasks, including zero-shot TTS, noise suppression using an optional transcript input, speech removal, target speaker extraction using an optional transcript input, and speech editing for both quiet and noisy acoustic environments (Fig. \ref{fig:SpeechX}). We refer to our proposed model as {\bf SpeechX}\footnote{X stands for transformation to highlight that our model performs various speech transformation tasks in addition to zero-shot TTS.}.
As with VALL-E, SpeechX adopts a language modeling approach that generates codes of a neural codec model, or acoustic tokens, based on textual and acoustic inputs. To enable the handling of diverse tasks, we incorporate additional tokens in a multi-task learning setup, where the tokens collectively specify the task to be executed.
Experimental results, using 60k hours of speech data from LibriLight~\cite{kahn2020libri} as a training set, demonstrate the efficacy of SpeechX, showcasing comparable or superior performance compared to expert models in all the aforementioned tasks. Notably, SpeechX also exhibits novel or expanded capabilities, such as preserving background sounds during speech editing and leveraging reference transcriptions for noise suppression and target speaker extraction.
From the perspective of speech enhancement tasks, SpeechX provides a simple and effective framework to leverage the reference transcription for improving intelligibility by integrating non-speech enhancement tasks, such as zero-shot TTS.
Audio samples showcasing the capabilities of our proposed SpeechX model are available at \url{https://aka.ms/speechx}.

\section{Related Work}
\subsection{Autoregressive generative models}
Generative models based on a language modeling approach using autoregressive Transformers, also known as decoder-only Transformers, have garnered significant success in various application domains. Notable examples of such models include the GPT series~\cite{NEURIPS2020_1457c0d6,openai2023gpt4} and DALL-E~\cite{pmlr-v139-ramesh21a}.
The autoregressive approach has also been extended to the audio and speech domains. AudioLM~\cite{borsos2023audiolm} and MusicLM~\cite{agostinelli2023musiclm} are pioneering efforts that exploit multiple types of tokens, each with a distinct time scale and degree of semantic granularity, allowing for hierarchical token generation. This hierarchical structure, comprising both coarse and fine-grained tokens, enables the synthesis of sounds with both nuanced details and long-term regularities. 

For zero-shot TTS, VALL-E~\cite{wang2023neural} and SPEAR-TTS~\cite{kharitonov2023speak} employ the autoregressive Transformers by representing textual (semantic) and acoustic tokens as a single data stream. This approach enables the models to perform zero-shot speaker adaptation, facilitating the generation of TTS voices that mimic a specific person's voice. It was demonstrated that these models could perform zero-shot TTS from speech clips as short as three seconds. 
A notable advantage of these autoregressive speech generation models is their ability to perform TTS without requiring a separate duration model. This streamlined architecture simplifies the training process and potentially offers increased flexibility needed to subsume various speech generation tasks. 
For this reason, we opt to build our SpeechX models by using autoregressive Transformers.

\subsection{Multi-task generative speech models}
Several papers have recently reported efforts in developing audio-text-based speech generation models that support zero-shot TTS and several related tasks. These tasks include voice or style conversion (Make-A-Voice~\cite{huang2023makeavoice}, NaturalSpeech2~\cite{shen2023naturalspeech}, and Voicebox~\cite{le2023voicebox}), speech editing (Mega-TTS~\cite{jiang2023mega} and Voicebox), and denoising (NaturalSpeech2 and Voicebox). 
Voicebox has showcased noteworthy advancements by facilitating a multitude of tasks through its masked speech prediction principle. Nevertheless, its capabilities are still limited to clean speech generation alone, falling short of effectively dealing with noisy speech or encompassing conventional audio enhancement tasks such as noise suppression and target speaker extraction. 

In this study, we deal with both clean and noisy speech and unify the generation and transformation tasks. To accomplish this, we extend VALL-E by performing multi-task learning with task-dependent prompts. The resulting model, which we call SpeechX, exhibits versatility in various speech processing tasks. The model excels not only in speech generation tasks like zero-shot TTS and speech editing but also performs effectively in enhancement tasks such as noise suppression and target speaker extraction. It also realizes novel capabilities, such as editing spoken content while retaining the background noise or effectively leveraging transcriptions for enhancement tasks.

\section{Method}

\subsection{Overview}
Fig.~\ref{fig:SpeechX} illustrates an overview of the SpeechX architecture. Building upon the principles introduced in VALL-E, SpeechX employs a neural codec language model based on Transformers.
The model learns to perform conditional generation of a neural code sequence, denoted as $\mathcal{O}$, based on two input prompts: textual prompt $\mathcal{T}$ and acoustic prompt $\mathcal{A}$. The neural codes may also be referred to as acoustic tokens. 

The textual prompt $\mathcal{T}$ is a sequence of phonemes obtained by applying grapheme-to-phoneme conversion\footnote{\url{https://github.com/Kyubyong/g2p}} to an input text. 
The textual prompt conveys the semantic information, and thus it is called semantic tokens. 
Conversely, the acoustic prompt $\mathcal{A}$ encapsulates the acoustic information of an input speech signal. 
It is obtained by converting the input audio into a sequence of acoustic tokens with an encoder of the neural codec model. 
Furthermore, to specify the task to be executed, or equivalently the desired output, we incorporate additional tokens in the acoustic prompt.
The details will be explained in Section \ref{sec:multi-task-training}.
The output $\mathcal{O}$ is a sequence of neural codes of the desired signal, which is then translated into a waveform signal with the codec decoder. 

We use EnCodec~\cite{defossez2022high} as the neural codec model, following the prior work. 
EnCodec is based on an encoder-decoder architecture with $L$ quantization layers. In our experiments, we use $L=8$ to be consistent with the configuration of \cite{wang2023neural}. Each layer of the EnCodec model produces discrete codes consisting of 1024 entries at a sampling rate of 75 Hz.

We emphasize that the proposed simple architecture capitalizes on the end-to-end modeling capability of the neural language modeling approach. In contrast to other zero-shot TTS or speech generation methods, this approach eliminates the need for a separate model, such as a speaker embedding model or a duration model, apart from the neural codec model. This key property allows SpeechX to acquire knowledge of diverse tasks with varying requirements and input-output relationships, thereby facilitating a versatile and highly extensible speech generation process.

\subsection{Neural codec language model}
\label{sec:design}
As with VALL-E~\cite{wang2023neural}, SpeechX makes use of auto-regressive (AR) and non-auto-regressive (NAR) Transformer models. 
Specifically, the AR model is 
used to output the neural codes corresponding to the first quantization layer of EnCodec. 
On the other hand, the NAR model generates the neural codes of all the layers above the first layer, namely the second through eighth layers. 
Combining the AR and NAR models provides a reasonable trade-off between generation flexibility and inference speed, as discussed in \cite{wang2023neural}.

Let output $\mathcal{O}$ be specifically represented as matrix $\mathbf{O} = [o_{t,l}] \in \mathbb{N}^{T\times L}$, where $o_{t,l}$ represents the code for the $l$-th codec layer at time frame $t$ and it can take one of the 1024 values. 
The output sequence length is denoted by $T$.
The AR model comprises a stack of Transformer decoder layers~\cite{NIPS2017_3f5ee243} and is optimized by minimizing the negative log-likelihood of the first layer code of the desired output, which is defined as follows:
\begin{align}
\mathcal{L_\textit{AR}} = -\sum_{t=1}^T{\log P(o_{t,1} | \mathcal{T}, \mathcal{A}, \mathbf{o}_{<t,1}; \theta_\textit{AR})},
\label{eq:ar}
\end{align}
where $\mathbf{o}_{<t,1} = [o_{1, 1}, \cdots, o_{t-1, 1}]$, while $\theta_\textit{AR}$ represents the AR Transformer model parameters. 
Different embedding projections are applied to the textual and acoustic tokens, and they are superimposed by sinusoidal positional embeddings. 

Note that the AR model in SpeechX is conditioned on three elements: the acoustic prompt $\mathcal{A}$, the textual prompt $\mathcal{T}$, and the past acoustic history $\mathbf{o}_{<t,1}$. This formulation differs from that of VALL-E, where the AR model is conditioned only on $\mathcal{T}$ and $\mathbf{o}_{<t,1}$ (Eq. (1) of \cite{wang2023neural}), and the audio prompt is represented as part of $\mathbf{o}_{<t,1}$ during inference. 
This difference provides a practical benefit during the inference time of zero-shot TTS, where SpeechX no longer requires a transcription of the audio prompt. More specifically, in the case of VALL-E inference, we need to construct $\mathcal{T}$ from the concatenation of the transcription of the audio prompt and the text prompt. The audio is then generated by setting the codec sequence of the audio prompt to $\mathbf{o}_{<t,1}$.
In contrast, for SpeechX inference, $\mathcal{T}$ is simply the text prompt. The model can generate the codec sequence without requiring the transcription of the audio prompt.

After obtaining the first layer codes with the AR model, 
the NAR model is used to generate the $l$-th layer codes based on the text and acoustic prompts as well as the output codes for the first $l-1$ layers, which have already been produced. 
The model is used repeatedly for $l = 2, \cdots, 8$. 
Since we use the same NAR model for the remaining seven layers, 
the NAR model is trained to minimize the following negative log-likelihood function: 
\begin{align}
\mathcal{L_\textit{NAR}} = -\sum_{l=2}^8{\log P(\mathbf{o}_{:,l} | \mathcal{T}, \mathcal{A}, \mathbf{o}_{:,<l}; \theta_\textit{NAR}  )},
\label{eq:nar}
\end{align}
where $\theta_\textit{NAR}$ represents the NAR model parameters, while 
$\bm{o}_{:, l}$ denotes the entire sequence of $o_{t, l}$ for the $l$th layer, and $\bm{o}_{:, <l} = [\bm{o}_{:, 1}, \cdots, \bm{o}_{:, l-1}]$.
In this formulation, in order for the single NAR model to process each of the seven layers, the acoustic tokens from the first to $(l-1)$th layers, $\textbf{o}_{:, <l}$, are embedded and summed up.

\subsection{Task-based prompting}
\label{sec:multi-task-training}

\begin{table*}[t!]
  \centering
  \caption{Task-based prompting: prompts and desired output for individual tasks. $\mathrm{G2P}(\cdot)$ denotes grapheme-to-phoneme conversion. }
  \tabcolsep = 1.5mm
  \label{tab:prompting}
 \resizebox{\textwidth}{!}{
  \begin{tabular}{@{}llllllll@{}}
  \toprule
  \multirow{1}{*}{\bf Task} && {\bf Textual prompt $\mathcal{T}$}  & {\bf Acoustic prompt $\mathcal{A}$} & {\bf Desired output $\mathcal{O}$} \\
  \midrule
  \multicolumn{1}{l}{Noise suppression}            && G2P(text) / null   & $\texttt{<ns>}$, $\mathrm{C}(s+n)$ & $\mathrm{C}(s)$ \\
  \multicolumn{1}{l}{Speech removal}               && G2P(text) / null   & \texttt{<sr>}, $\mathrm{C}(s+n)$ & $\mathrm{C}(n)$ \\
  \multicolumn{1}{l}{Target speaker extraction}    &&  G2P(text) / null   &  $\mathrm{C}(s'_1)$, \texttt{<tse>}, $\mathrm{C}(s_1+s_2)$ & $\mathrm{C}(s_1)$ \\
  \multicolumn{1}{l}{Zero-shot TTS}                && G2P(text)  &  $\mathrm{C}(s)$  & $\mathrm{C}(s')$ \\
  \multicolumn{1}{l}{Clean speech editing}         && G2P(text)  &  $\mathrm{C}(s_{\rm pre})$, \texttt{<soe>}, \texttt{<mask>}, \texttt{<eoe>}, $\mathrm{C}(s_{\rm post})$ & $\mathrm{C}(s_{\rm pre}), \mathrm{C}(s_{\rm edit}), \mathrm{C}(s_{\rm post})$\\
  \multicolumn{1}{l}{Noisy speech editing}         && G2P(text)  &  $\mathrm{C}(s_{\rm pre}+n_{\rm pre})$, \texttt{<soe>}, $\mathrm{C}(s_{\rm mid}+n_{\rm mid})$, \texttt{<eoe>}, $\mathrm{C}(s_{\rm post}+n_{\rm post})$ & $\mathrm{C}(s_{\rm pre}+n_{\rm pre}), \mathrm{C}(s_{\rm edit}+n_{\rm mid}), \mathrm{C}(s_{\rm post}+n_{\rm post})$\\
  \bottomrule 
\end{tabular}
}
\end{table*}

\label{sec:prompting}

SpeechX aims to handle multiple tasks with one model.
To this end, we adopt task-based prompting, as illustrated
in Table \ref{tab:prompting} and explained in detail below.

\textbf{Noise suppression} is a task of extracting clean speech signal $s$ from its noise-corrupted observation $s+n$, where $n$ denotes the noise. 
For the noise suppression task, we incorporate a special token, denoted as $\texttt{<ns>}$, 
to form the acoustic prompt, resulting in $\mathcal{A}=[\texttt{<ns>}, \mathrm{C}(s+n)]$. Here, $\mathrm{C}(\cdot)$ denotes the function used to convert an audio signal into a neural codec token sequence. 
While the textual prompt $\mathcal{T}$ is supposed to be provided by a user as a reference transcription, 
we let the use of the textual prompt be optional to accommodate the scenario where the human transcription is unavailable.
The desired output is the acoustic token sequence of the clean audio, $\mathrm{C}(s)$.

\textbf{Speech removal} involves removing speech from a noisy speech signal while preserving the background noise. It is useful for removing only unwanted speech from recordings.
To address this task, we employ a special token, $\texttt{<sr>}$,  
to construct the acoustic prompt as $\mathcal{A}=[\texttt{<sr>}, \mathrm{C}(s+n)]$. The desired output is the acoustic token sequence of the noise signal, $\mathrm{C}(n)$.
As in the case of noise suppression, the textual prompt can be omitted.

\textbf{Target speaker extraction} aims at isolating clean speech $s_1$ of a target speaker from a mixture of
$s_1$ and interfering speech $s_2$ from a secondary speaker. The target speaker is identified through a short enrollment audio $s'_1$ of that individual, where we assumed three seconds for the enrollment. For this task, 
we form the acoustic prompt by concatenating the acoustic tokens extracted from the enrollment audio, $\mathrm{C}(s'_1)$, and those of the mixed speech, $\mathrm{C}(s_1+s_2)$, with a task-specifying token, denoted as $\texttt{<tse>}$. That is, 
we have $\mathcal{A}=[\mathrm{C}(s'_1), \texttt{<tse>}, \mathrm{C}(s_1+s_2)]$.
The desired output is $\mathrm{C}(s_1)$.
As with the previous tasks, the inclusion of the textual prompt is optional. 

\textbf{Zero-shot TTS} aims to generate a speech signal $s'$ by leveraging both the provided input text and an enrollment speech $s$. The goal is to ensure that the speech characteristics of $s'$ closely resemble those of $s$, while also accurately reflecting the input text. For this task, we employ the acoustic tokens extracted from the enrollment audio, denoted as $\mathrm{C}(s)$, as the acoustic prompt. 
The model generates acoustic tokens for the synthesized speech, $\mathrm{C}(s')$, based on the input text. These acoustic tokens are then converted into the corresponding waveform.

\textbf{Clean speech editing} is defined as modifying a segment of input speech to align with an input text. 
Let $s$ denote the input speech signal to be edited.
We divide $s$ into three distinct portions, $s_{\rm pre}$, 
$s_{\rm mid}$, and $s_{\rm post}$,
with $s_{\rm mid}$ being the target segment for editing, without loss of generality ($s_{\rm pre}$ and $s_{\rm post}$ can be empty). 
We construct the acoustic prompt as 
$[\mathrm{C}(s_{\rm pre}), \texttt{<soe>}, \texttt{<mask>}, \texttt{<eoe>}, \mathrm{C}(s_{\rm post})]$,
where new tokens \texttt{<soe>}, \texttt{<mask>}, \texttt{<eoe>}
are introduced to specify the task and the speech segment designated for editing. 
The desired output is a sequence of 
neural codes, $[\mathrm{C}(s_{\rm pre}), \mathrm{C}(s_{\rm edit}),\mathrm{C}(s_{\rm post})]$, where the spoken content of $[  s_{\rm pre}, s_{\rm edit},  s_{\rm post} ]$ matches the input text. The speaker characteristics of $s_{\rm edit}$ must be consistent with those of $s_{\rm pre}$  and $s_{\rm post}$.

\textbf{Noisy speech editing}, in contrast, operates on noisy speech as input, aiming to modify the speech content within a segment while keeping the underlying background noise intact. 
Therefore, this task would be more challenging than
the clean speech editing task because the model
needs to distinguish between speech and noise during the editing process. 
To accomplish this objective, it is crucial to provide the model with the complete input speech signal instead of masking out the segment for editing with a $\texttt{<mask>}$ token. 
Therefore, 
we construct the acoustic prompt as 
$[\mathrm{C}(s_{\rm pre}+n_{\rm pre}), \texttt{<soe>}, \mathrm{C}(s_{\rm mid}+n_{\rm mid}), \texttt{<eoe>}, \mathrm{C}(s_{\rm post}+n_{\rm post})]$, with the subscripts corresponding to pre, mid, or post as previously defined.
The desired output comprises a sequence of neural codes,
$[\mathrm{C}(s_{\rm pre}+n_{\rm pre}), \mathrm{C}(s_{\rm edit}+n_{\rm mid}), \mathrm{C}(s_{\rm post}+n_{\rm post})]$. 
This formulation makes it clear that  the model must transform $s_{\rm mid}$ into $s_{\rm edit}$ based on the text input while retaining $n_{\rm mid}$.

In practical speech editing scenarios, the input text is often obtained by first applying automatic speech recognition (ASR) to the input speech and then having a user edit the transcription. In such situations, it is simple to identify the positions at which $\texttt{<soe>}$ and $\texttt{<eoe>}$ must be inserted. 
Also, it is noteworthy that, in clean speech editing,  the use of $\texttt{<mask>}$ allows the model to adaptively change the output speech length in such a way that the output speech sounds natural in terms of speaking speed. 

The outlined task-based prompting strategy equips the SpeechX model with the ability to uniquely decide the desired output during inference. This approach enables flexibility for incorporating additional tasks. Adding new tasks entails integrating corresponding prompting schemes and continuing model training from an existing checkpoint,
where only embeddings for newly introduced task-specific tokens are randomly initialized. This can be performed without changing the underlying model architecture.

\subsection{Model training}
\label{sec:training}

During training, we randomly sample the task for each model update at an equal probability.  This is intended to ensure the model does not unduly favor any particular tasks. For noise suppression, speech removal, and target speaker extraction tasks, we include the textual prompt at a 50\% probability so that the model equally experiences both text and text-less scenarios. 

To help the model to acquire basic generation capabilities, we first train the model only for zero-shot TTS and then continue the training process using all the tasks to perform multi-task learning. 
In other words, we initialize the model with an existing VALL-E model checkpoint.
Precisely speaking, the SpeechX model trained solely for zero-shot TTS exhibits slight divergence from VALL-E. This difference arises from the fact that the former explicitly incorporates a distinct enrollment audio, originating from the same speaker, for each training sample, while the latter does not. Nevertheless, for the sake of simplicity, we refer to this initialization approach as VALL-E initialization.
When starting the multi-task training stage, randomly initialized embeddings are appended for the special tokens related to the task-dependent prompts. 
This two-stage training strategy substantially enhances performance across all tasks, as evidenced by our experimental results.

\section{Evaluation Setups}

Evaluating versatile speech generation models like SpeechX requires performing an array of tests, each focusing on individual tasks. To keep the experiments manageable as well as ensure consistency across the tasks, we used evaluation datasets that were derived from the test-clean split of LibriSpeech for all evaluations. In this section, we provide the details of our evaluation setups. Following previously established practices~\cite{le2023voicebox,wang2023neural}, we selected the test samples with durations between 4 and 10 seconds. 

\subsection{Evaluation data}

\textbf{Zero-shot TTS:} For each test sample, we used the reference transcription to create the textual prompt. The acoustic prompt was generated by randomly choosing another utterance of the same speaker and extracting a 3-second-long clip. 
\vspace{.5em}

\textbf{Noise suppression:} We mixed each test sample with a noise sample randomly picked from the MUSAN dataset~\cite{snyder2015musan} at a signal-to-noise ratio (SNR) which was randomly determined from the range between 0 dB and 20 dB. The task was to recover the uncorrupted speech from the noisy speech. The acoustic prompt was obtained by applying EnCodec to the noisy signal. Regarding the textual prompt, we considered both text-less (i.e., using no semantic prompt) and text-guided noise suppression, where we used the reference transcription for the text-guided setting. 
\vspace{.5em}

\textbf{Target speaker extraction:} We mixed each test sample with an utterance of a different speaker at a signal-to-interference ratio (SIR) which was randomly determined from the range between 0 dB and 20 dB. Also, we randomly chose one or more other utterances of the same speaker to create a 3-second-long enrollment clip to help models identify who the desired speaker is. Both the mixed and enrollment signals were used to derive the acoustic prompt as described in Section \ref{sec:multi-task-training}. The task was to recover the original uncorrupted speech of the target speaker. As with the noise suppression task, we considered both text-less and text-guided settings. 
\vspace{.5em}

\textbf{Clean speech editing:} For each test sample, we randomly selected a period of length between 10\% and 50\% of the whole utterance. We replaced the speech of the selected period with another randomly chosen speech sample of the same speaker. Given the partially replaced, speaker homogeneous speech and the reference transcription, the task was to generate a speech signal that follows the transcription without changing the speaker characteristics and the unreplaced portion of the input signal. 
In our experiments, we used the correct $\texttt{<soe>}$ and $\texttt{<eoe>}$ locations based on the knowledge of the replaced segment. 
\vspace{.5em}

\textbf{Noisy speech editing:} We added a randomly picked MUSAN noise sample to each test sample of the clean speech editing task. The SNR was chosen from the range of 0 dB to 20 dB. Given the noise-corrupted partially replaced speech and the reference transcription, the task was to generate a noisy speech signal that follows the transcription without changing the background noise, the speaker characteristics, and the unreplaced portion of the input speech. 
\vspace{.5em}

\textbf{Speech removal:} The same dataset was used as the one used for noise suppression. Given a noisy speech signal, the task was to extract the noise signal by removing the speech. We considered only the textless case. 
Consequently, the input exclusively comprised the acoustic prompt corresponding to the noisy speech.

\subsection{Metrics}

For consistency and reproducibility, we opted to use objective metrics for individual tasks as described below. 
\vspace{.5em}

\textbf{Word error rate (WER):} We employed the WER as a metric to evaluate the fidelity of the generated audio in adhering to the provided transcription. The ASR system utilized for our experiments was NeMo's stt\_en\_conformer\_transducer\_large model\footnote{\url{https://huggingface.co/nvidia/stt\_en\_conformer\_transducer\_xlarge}}, which is based on the Conformer Transducer architecture~\cite{gulati20_interspeech}. We selected this particular ASR model based on its superior stability and robustness against noise and processing artifacts in comparison to other publicly available ASR models, as was observed during our preliminary experiments. Robustness in ASR is particularly crucial for tasks such as noise suppression and noisy speech editing.
The WER metric was employed across all tasks, with the exception of speech removal.

\vspace{.5em}

\textbf{Speaker similarity score (SIM):} The speaker similarity score served as a metric to assess the coherence of the generated speech in relation to the speaker's characteristics. This score was calculated as the cosine similarity between the speaker embeddings of the generated speech and the desired speech signals. The computation of speaker embeddings was performed using NeMo's TitaNet-Large\footnote{\url{https://huggingface.co/nvidia/speakerverification\_en\_titanet\_large}}. 
We employed the original audio data instead of utilizing an EnCodec-processed signal for the speaker similarity measurement to capture and reflect any potential speech deformation effects that may arise due to the use of the EnCodec model. SIM was used in zero-shot TTS, clean speech editing, and noisy speech editing. 
\vspace{.5em}

\textbf{DNSMOS:} For evaluation in the noise suppression and target speaker extraction tasks, we utilized DNSMOS~\cite{reddy2021dnsmos}, a well-established model-based metric for predicting the perceived quality of acoustically corrupted speech\footnote{\url{https://github.com/microsoft/DNS-Challenge/tree/master/DNSMOS}}. Specifically, we employed the OVRL score from the DNSMOS P.835 model. To evaluate the performance of target speaker extraction, we employed a personalized DNSMOS model, which was tailored for this particular task and is available on the same webpage.
\vspace{.5em}

\textbf{Perceptual Evaluation of Speech Quality (PESQ):} For the noise suppression and target speaker extraction tasks, we also utilized PESQ~\cite{rix2001perceptual}. Unlike DNSMOS, PESQ is an intrusive metric that necessitates the clean reference signals. Consequently, PESQ is expected to assess the fidelity of the generated audio with respect to the original clean data.
\vspace{.5em}

\textbf{Mel-cepstral distortion (MCD):} MCD\footnote{\url{https://pypi.org/project/pymcd}} is a metric used to quantify the dissimilarity between two sequences of mel cepstra. We employed this metric to objectively measure the speech removal accuracy by comparing the estimated noise with the ground truth noise audio.

\section{Experiments}
\subsection{Training data}

We sourced clean speech data from LibriLight, comprising 60 thousand hours of untranscribed English reading speech from over 7,000 speakers~\cite{kahn2020libri}, as was performed in the zero-shot TTS experiment using VALL-E~\cite{wang2023neural}.
To meet the specific training requirements for each task, data simulation was performed by following the methods employed for creating the evaluation data, as elaborated below. Note that, as discussed in Section \ref{sec:training}, we formed individual training mini-batches based on randomly selected tasks for each iteration.

For the noise suppression and speech removal tasks, we mixed the clean speech with noise samples from the DNS challenge corpus~\cite{dubey2023icassp} at SNRs between -5 dB and 20 dB. Our models were trained to recover the acoustic tokens of the clean speech and noise for noise suppression and speech removal, respectively. 
For the target speaker extraction task, we mixed the individual clean speech samples with those of other randomly chosen speakers with SIRs ranging from -5 dB to 20 dB.
Regarding clean speech editing, for each clean utterance, we randomly selected a subsegment of length ranging from 10\% to 70\%, and then substituted it with another audio segment from the same speaker with different content. We saved the start and end times of the replaced segment, which were used to insert the $\texttt{<soe>}$ and $\texttt{<eoe>}$ tokens to the correct positions in the acoustic prompt during training. 
Furthermore, to create training samples for noisy speech editing, we added noise samples used in the noise suppression task to the partially replaced clean audio. As a result, we obtained pairs of noisy partially replaced speech and the corresponding original noisy speech, which served as the training data for the noisy speech editing task. The SNR range used for noisy speech editing training was also from $-5$ dB to $20$ dB.

Since LibriLight does not provide reference transcriptions, we adopted a pseudo-labeling approach to derive the semantic prompts, i.e., the phoneme sequences of the individual training samples, by following \cite{le2023voicebox,wang2023neural}.
Specifically, we transcribed the LibriLight training data with an off-the-shelf Kaldi model that was trained on the 960-hour Librispeech data with 3x speed perturbation\footnote{\url{https://kaldi-asr.org/models/m13}}.

\subsection{Model and training configurations}

Both the SpeechX AR and NAR models share the same Transformer architecture, featuring 12 layers, 16 attention heads, an embedding dimension of 1024, a feed-forward layer dimension of 4096, and a dropout rate of 0.1. 

We conducted experiments employing two initialization methods: random initialization and VALL-E initialization (refer to Section~\ref{sec:training} for details). In the random initialization scenario, we trained the SpeechX model for 800k iterations. The model optimization utilized the AdamW optimizer, with the learning rate undergoing a warm-up phase for the initial 32K updates, peaking at $5\times10^{-4}$, before transitioning into a linear decay phase. Conversely, with VALL-E initialization, we opted for 400k iterations, as the initial model already underwent zero-shot TTS training over 400k iterations. In this instance, the learning rate scheduler was retained, but the warm-up period was shortened to the first 20k updates.

\subsection{Baseline expert models}
We employed expert models for different tasks to establish comparison baselines. For zero-shot TTS, we utilized VALL-E by following the model configuration outlined in the original paper~\cite{wang2023neural}. For the noise suppression task, we employed a non-causal Deep Complex Convolutional Recurrent Network (DCCRN)~\cite{hu2020dccrn}, which is a widely recognized model for noise suppression. Our training data for DCCRN came from Microsoft's internal dataset, and we further fine-tuned the model using the ASR objective based on the training recipe of \cite{eskimez2021human}.
For target speaker extraction, we leveraged VoiceFilter~\cite{wang2018voicefilter}, employing a bidirectional LSTM configuration. We relied on a publicly available implementation of VoiceFilter\footnote{\url{https://github.com/Edresson/VoiceSplit}}.
Finally, for speech editing, we employed A$^3$T~\cite{bai20223} as the baseline. The implementation of A$^3$T that we used is also publicly accessible\footnote{\url{https://github.com/richardbaihe/a3t}}.

\begin{table*}[!t]
    \caption{
    Results
    for various speech generation/transformation tasks by SpeechX compared to expert models for individual tasks. Textual prompts were used for noise suppression and target speaker extraction. In zero-shot TTS, ``no processing'' row shows the results of desired speech signals.}
  \label{tab:results}
  \tabcolsep = 0.9mm
  \centering
 \resizebox{\textwidth}{!}{
\begin{tabular}{llccccccccccccccccccc}
  \toprule
\multicolumn{1}{c}{\multirow{2}{*}{\bf Model}}  && 
\multicolumn{3}{c}{\bf Noise suppression} && 
\multicolumn{3}{c}{\bf Target speaker extraction} && 
\multicolumn{2}{c}{\bf Zero-shot TTS} && 
\multicolumn{2}{c}{\bf Clean speech editing} &&
\multicolumn{2}{c}{\bf Noisy speech editing}  &&
\multicolumn{1}{c}{\bf Speech removal} 
\\ 
\cmidrule{3-5} \cmidrule{7-9} \cmidrule{11-12} \cmidrule{14-15} \cmidrule{17-18} \cmidrule{20-20}
\multicolumn{1}{c}{}  &  & 
\multicolumn{1}{c}{WER$\downarrow$} & \multicolumn{1}{c}{DNSMOS$\uparrow$} & \multicolumn{1}{c}{PESQ$\uparrow$} & & 
\multicolumn{1}{c}{WER$\downarrow$} & \multicolumn{1}{c}{DNSMOS$\uparrow$} & \multicolumn{1}{c}{PESQ$\uparrow$} & & 
\multicolumn{1}{c}{WER$\downarrow$} & \multicolumn{1}{c}{SIM$\uparrow$} & &
\multicolumn{1}{c}{WER$\downarrow$} &  \multicolumn{1}{c}{SIM$\uparrow$} &&
\multicolumn{1}{c}{ WER$\downarrow$}  & \multicolumn{1}{c}{ SIM$\uparrow$}   &&
\multicolumn{1}{c}{MCD$\downarrow$} 
\\ 
\toprule
\bf No processing             && 
3.29  & 2.42  & 1.93 && 
12.55 & 3.04  & 2.27 && 
1.71 & 1.00 &&  
38.29 & 0.96  && 
42.48 & 0.87 &&  
12.57 \\
\midrule
\multicolumn{1}{l}{\multirow{2}{*}{\bf Expert model}} & & 
\multicolumn{3}{c}{\bf DCCRN~\cite{hu2020dccrn, eskimez2021human}} & &
\multicolumn{3}{c}{\bf  VoiceFilter~\cite{wang2018voicefilter}} & & 
\multicolumn{2}{c}{\bf  VALL-E~\cite{wang2023neural}} & & 
\multicolumn{2}{c}{\bf  A$^3$T~\cite{bai20223}} &&
\multicolumn{2}{c}{\bf  A$^3$T~\cite{bai20223}} &&
\multicolumn{1}{c}{\multirow{2}{*}{\bf N/A}}\\
\cmidrule{3-5} \cmidrule{7-9} \cmidrule{11-12} \cmidrule{14-15} \cmidrule{17-19}
\multicolumn{1}{c}{}             &&  
6.39     &  3.25    &  3.52    &&      
5.09   &  3.39 &  2.90 && 
5.90   & 0.57 &&
17.17 & 0.29 &&  
32.17 & 0.18 &&
\\ 

\midrule
\bf  SpeechX (random init.)  && 
2.56 & 3.05 & 2.24 && 
3.12 & 3.46 & 2.27 && 
5.40 & 0.57 &&
8.10 & 0.75 &&
15.33 & 0.64 &&
3.04
\\
\bf SpeechX (VALL-E init.)  &&  
2.48 & 3.05 & 2.24 && 
2.53 & 3.46  &  2.28 &&  
4.66 & 0.58 &&  
5.63 & 0.76 && 
13.95 & 0.65 &&
3.05  \\
\bottomrule 
\end{tabular}
}
\end{table*}

\begin{table}[t]
  \centering
  \caption{Results of noise suppression and target speaker extraction with or without textual prompt.}
  \tabcolsep = 0.9mm
  \label{tab:text_ns}
{\footnotesize
  \begin{tabular}{@{}ccccccccc@{}}
  \toprule
   \multirow{2}{*}{Prompt} && \multicolumn{3}{c}{Noise suppression}   && \multicolumn{3}{c}{Target speaker extraction} \\ 
    \cmidrule{3-5} \cmidrule{7-9}
     && { WER$\downarrow$}    & { DNSMOS$\uparrow$} & { PESQ$\uparrow$} && { WER$\downarrow$}  & { DNSMOS$\uparrow$} & { PESQ$\uparrow$} \\
  \midrule
  w/ text    && 2.48 & 3.05 & 2.24 && 2.53 & 3.46 & 2.28 \\
  w/o text   && 6.76 & 3.05 & 2.20  && 5.00 & 3.01 & 2.23 \\ 
  \bottomrule 
\end{tabular}
}
\end{table}

\subsection{Results}

\subsubsection{Result overview}
Table.~\ref{tab:results} shows the performance analysis of SpeechX in various tasks compared to the individual expert models.  We can see that initializing the model parameters using an exiting VALL-E model checkpoint was beneficial across all tasks, especially in terms of WER. 

In noise suppression and target speaker extraction, SpeechX exhibited superior performance in terms of WER compared to the respective expert models. Conventional regression-based noise suppression and target speaker extraction models are known to suffer from processing artifacts, which our WER results confirmed. SpeechX was able to avoid this detrimental effect thanks to the audio-text-based generation capability. 
On the other hand, in terms of DNSMOS and PESQ scores, it lagged behind the expert models. This can largely be attributed to the impact of the codec model used, as discussed in detail in Section \ref{sec:codec}. 
The investigation into the speech removal task revealed that SpeechX demonstrated substantial improvement in MCD, showcasing its efficacy in removing speech. 
These results underscore the versatility of the SpeechX model in handling enhancement-related tasks, while also highlighting the usefulness of the audio-text-based speech generation capability that SpeechX provides.

In the zero-shot TTS task, SpeechX demonstrated a slight advantage over the baseline VALL-E model in terms of WER while concurrently achieving a comparable speaker similarity score\footnote{To avoid potential confusion, it should be noted that our experimental setup corresponds to the non-continual evaluation configuration utilized in the original VALL-E work.}. Furthermore, for the clean speech editing task, SpeechX exhibited significant improvement over the baseline A$^3$T model. 
The WER observed in the speech editing task was slightly higher than the WER obtained in the zero-shot TTS task, even though one might anticipate that they should fall within the same range. This discrepancy could be attributed to certain test samples where the length of non-edited speech was shorter than three seconds.
These results highlight that SpeechX is equally effective in tasks primarily focusing on speech generation capability, rather than transformation ability.

\begin{table*}[t]
    \caption{Impact of multi-task training. For models with random initialization, we trained a model for 800k iterations. For models with VALL-E initialization, the model was first trained with the zero-shot TTS task for 400k iterations, and then the model was further updated with various tasks for another 400k iterations.
    ZS: zero-shot, CSE: clean speech editing, NSE: noisy speech editing, NS: noise suppression, SR: speech removal, TSE: target speaker extraction.}
    \label{tab:multi-task}
  \centering
 \resizebox{\textwidth}{!}
{
\begin{tabular}{@{}cllcccccccccccccccc@{}}
  \toprule
\multirow{2}{*}{\bf Init.} &
\multicolumn{1}{c}{\multirow{2}{*}{\bf Training tasks}}  && 
\multicolumn{2}{c}{\bf Zero-shot TTS} && 
\multicolumn{2}{c}{\bf Clean speech editing}  && 
\multicolumn{2}{c}{\bf Noisy speech editing}  && 
\multicolumn{2}{c}{\bf Noise suppression} && 
\multicolumn{1}{c}{\bf Speech removal} &&
\multicolumn{2}{c}{\bf Target speaker extraction}   
\\
&
\multicolumn{1}{c}{}  && 
\multicolumn{1}{c}{WER$\downarrow$} & \multicolumn{1}{c}{SIM$\uparrow$}  &&  
\multicolumn{1}{c}{WER$\downarrow$} & \multicolumn{1}{c}{SIM$\uparrow$} && 
\multicolumn{1}{c}{WER$\downarrow$} & \multicolumn{1}{c}{SIM$\uparrow$} && 
\multicolumn{1}{c}{WER$\downarrow$} & \multicolumn{1}{c}{DNSMOS$\uparrow$} &&
\multicolumn{1}{c}{MCD$\downarrow$} &&
\multicolumn{1}{c}{WER$\downarrow$} & \multicolumn{1}{c}{DNSMOS$\uparrow$} 
\\ 
\toprule
\bf Rand. & \bf ZS-TTS                              && 5.90 & 0.57 &&  -    &   -  &&   -     &   -     &&         &       && - && - & - \\
\bf Rand. & \bf ZS-TTS + CSE + NSE + NS + SR + TSE  && 5.40 & 0.57 &&  8.10 & 0.75 && 15.33   &   0.64  &&   2.56  &  3.05 && 3.04 && 3.12 & 3.46 \\
\hdashline[1pt/2pt]\hdashline[0pt/1pt] 
{\bf VALL-E} & \bf CSE  && - & - && 5.57 & 0.76 && -  &  - &&   -  &   -   && -    && - & -\\
{\bf VALL-E} & \bf NSE  && - & - && - & - && 12.18  &  0.65 &&   -  &   -   && -    && - & -\\
{\bf VALL-E} & \bf NS  && - & - && -    &  -   &&  -     &    -   && 4.21&  3.04 && -    && - & -\\
{\bf VALL-E} & \bf SR  && - & - && -    &  -   &&   -    &   -    && -   & -     && 3.04 && - & -\\
{\bf VALL-E} & \bf TSE && - & - && -    &  -   &&   -    &   -    && -   & -     && -    && 3.97 & 3.51\\
{\bf VALL-E} &\bf ZS-TTS + CSE + NSE + NS + SR + TSE    && 4.66 & 0.58 && 5.63 & 0.76 && 13.95  & 0.65 &&  2.48 & 3.05 && 3.05 &&  2.53 & 3.46 \\
\bottomrule
\end{tabular}
}
\end{table*}

\subsubsection{Speech editing for clean and noisy speech}

Table \ref{tab:results} also compares the speech editing results between clean and noisy speech in terms of WER and SIM. Editing noisy speech poses greater challenges than clean speech, as it requires modifying the spoken content while preserving background noise. This difficulty is evident from a  WER gap of 38.29\% vs. 42.48\% observed between the clean and noisy audio signals to be edited as well as A$^3$T's limited WER improvement from 42.48\% to 32.17\%.

\begin{figure}
    \centering
    \includegraphics[width=\columnwidth,clip]{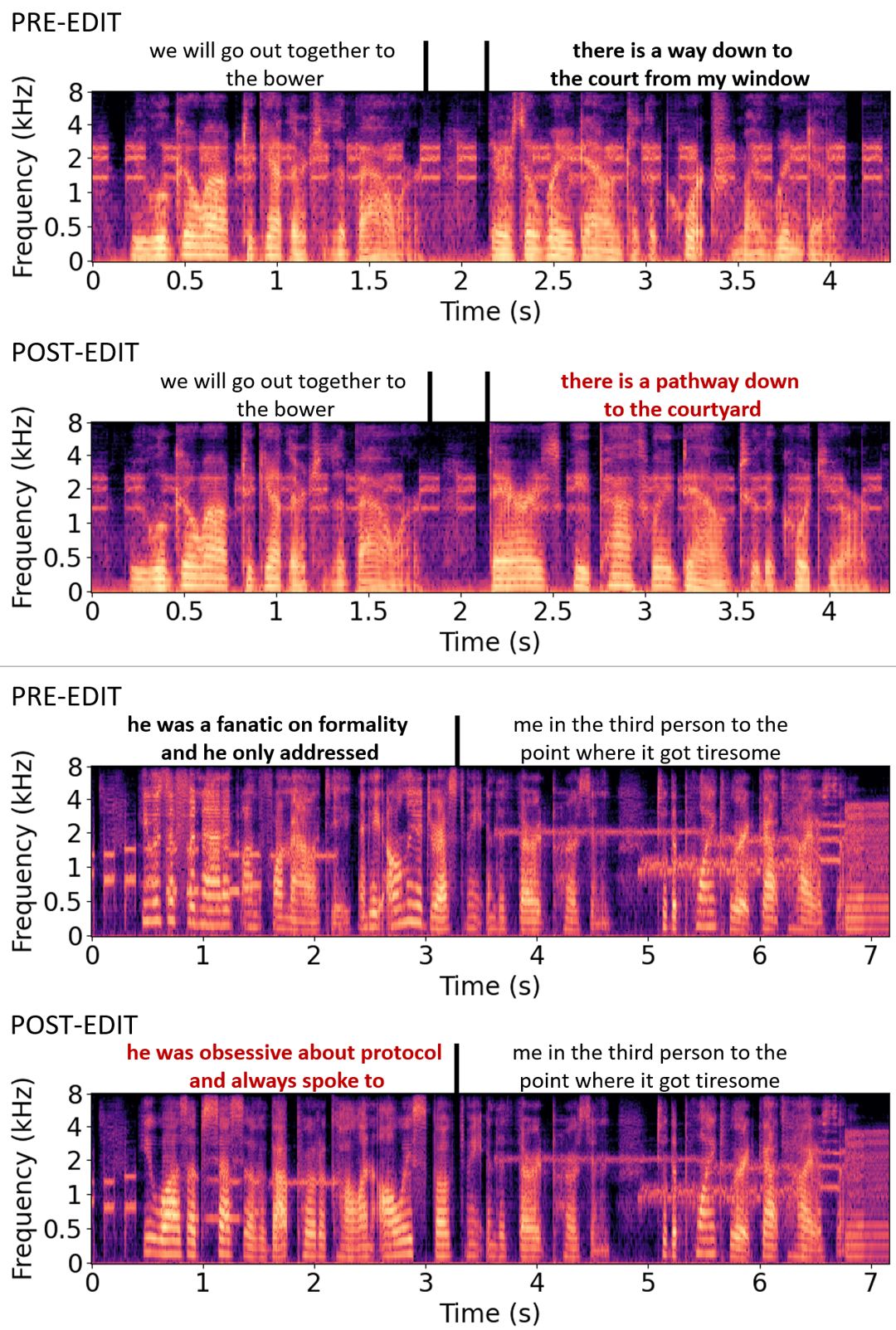}
    \caption{Mel spectrograms of pre-edit and post-edit noisy signals. The pre-edit signal was obtained by applying EnCodec compression and decompression without intermediate processing to highlight the change made by SpeechX's neural codec language model. See Section~\ref{sec:codec} for discussion on codec's impact.}
    \label{fig:specgram}
\end{figure}

Nonetheless, the SpeechX model successfully edited the noisy speech, reducing the WER to 13.95\% after processing. This demonstrates the model's robustness to acoustic noise in the input signal. The high SIM score of 0.65 shows the model largely preserved speaker characteristics, even with noise present. Our observation revealed the model retained background noise, as confirmed by our provided demo samples. Fig.~\ref{fig:specgram} compares mel spectrograms for two exemplary pairs of input and generated speech signals. In the first example, the input speech contained periodic noise in the middle frequency range. SpeechX preserved this background noise over the full input duration while selectively modifying only the foreground speech during the period beginning at two seconds. A similar observation can be made for the second example, wherein the alteration was applied to the first half of the speech content. In summary, the results demonstrate SpeechX model's effectiveness at noisy speech editing while maintaining speaker identity and background noise. Future work should develop a metric to quantitatively evaluate noise cloning capability.

\begin{table*}[t]
    \caption{Effects of adding tasks during training. ZS: zero-shot, SE: speech Editing, NS: noise suppression, SR: speech removal, TSE: target speaker extraction.}
    \label{tab:multi-task-effect}
  \centering
 \resizebox{\textwidth}{!}
{
\begin{tabular}{llccccccccccccc}
  \toprule
\multicolumn{1}{c}{\multirow{2}{*}{\bf Training tasks}}  && 
\multicolumn{2}{c}{\bf Zero-shot TTS} && 
\multicolumn{2}{c}{\bf Speech editing (clean/noisy)}  && 
\multicolumn{2}{c}{\bf Noise suppression} && 
\multicolumn{1}{c}{\bf Speech removal} &&
\multicolumn{2}{c}{\bf Target speaker extraction}   
\\
\multicolumn{1}{c}{}  && 
\multicolumn{1}{c}{WER$\downarrow$} & \multicolumn{1}{c}{SIM$\uparrow$}  &&  
\multicolumn{1}{c}{WER$\downarrow$} & \multicolumn{1}{c}{SIM$\uparrow$} && 
\multicolumn{1}{c}{WER$\downarrow$} & \multicolumn{1}{c}{DNSMOS$\uparrow$} &&
\multicolumn{1}{c}{MCD$\downarrow$} &&
\multicolumn{1}{c}{WER$\downarrow$} & \multicolumn{1}{c}{DNSMOS$\uparrow$} 
\\ 
\toprule
\bf ZS-TTS          && 5.90 & 0.57 &&  -   &    -  &&   -  &   - && - && - & - \\
\bf ZS-TTS + SE                  && 4.55 & 0.58 && 5.79\quad/\quad13.80  &  0.76\quad/\quad0.65  &&   -  &   - && - && - & -\\
\bf ZS-TTS + SE + NS/SR          && 5.11 & 0.57 && 6.91\quad/\quad 13.23 & 0.77\quad/\quad0.66 && 2.59 & 3.03 && 3.04 && - & -\\
\bf ZS-TTS + SE + NS/SR + TSE    && 4.66 & 0.58 && 5.63\quad/\quad13.95  & 0.76\quad/\quad0.65 &&  2.48 & 3.05 && 3.05 &&  2.53 & 3.46 \\
\bottomrule
\end{tabular}
}
\end{table*}

\begin{table*}[!t]
    \caption{Comparison between BPE-based and phoneme-based models.}
  \label{tab:bpe}
  \tabcolsep = 0.9mm
  \centering
 \resizebox{\textwidth}{!}{
\begin{tabular}{@{}llccccccccccccccccccc@{}}
  \toprule
\multicolumn{1}{c}{\multirow{2}{*}{\bf Model}}  && 
\multicolumn{3}{c}{\bf Noise suppression} && 
\multicolumn{3}{c}{\bf Target speaker extraction} && 
\multicolumn{2}{c}{\bf Zero-shot TTS} && 
\multicolumn{2}{c}{\bf Clean speech editing} &&
\multicolumn{2}{c}{\bf Noisy speech editing}  &&
\multicolumn{1}{c}{\bf Speech removal} 
\\ 
\cmidrule{3-5} \cmidrule{7-9} \cmidrule{11-12} \cmidrule{14-15} \cmidrule{17-18} \cmidrule{20-20}
\multicolumn{1}{c}{}  &  & 
\multicolumn{1}{c}{WER$\downarrow$} & \multicolumn{1}{c}{DNSMOS$\uparrow$} & \multicolumn{1}{c}{PESQ$\uparrow$} & & 
\multicolumn{1}{c}{WER$\downarrow$} & \multicolumn{1}{c}{DNSMOS$\uparrow$} & \multicolumn{1}{c}{PESQ$\uparrow$} & & 
\multicolumn{1}{c}{WER$\downarrow$} & \multicolumn{1}{c}{SIM$\uparrow$} &&
\multicolumn{1}{c}{WER$\downarrow$} & \multicolumn{1}{c}{SIM$\uparrow$} &&
\multicolumn{1}{c}{WER$\downarrow$} & \multicolumn{1}{c}{SIM$\uparrow$} &&
\multicolumn{1}{c}{MCD$\downarrow$} 
\\ 
\toprule
\bf  BPE && 
2.37 & 3.06 & 2.25 && 
2.46 & 3.53 & 2.32 && 
8.25 & 0.53 &&
7.44 & 0.76 &&
13.70 & 0.65 &&
3.06
\\
\bf Phoneme &&  
2.48 & 3.05 & 2.24 && 
2.53 & 3.46  &  2.28 &&  
4.66 & 0.58 &&  
5.63 & 0.76 && 
13.95 & 0.65 &&
3.05  \\
\bottomrule 
\end{tabular}
}
\end{table*}

\subsubsection{Effectiveness of text input in noise suppression and target speaker extraction}
\label{sec:notext}
With SpeechX, it is feasible to perform noise suppression and target speaker extraction using solely the acoustic prompt as input. To assess the efficacy of incorporating additional text input in the SpeechX model, we conducted noise suppression and target speaker extraction experiments where we employed only the acoustic prompt as the model input. Specifically, the input for noise suppression comprised the noisy speech, while for target speaker extraction, it consisted of the mixed speech and the target speaker's enrollment audio.

The experimental results are presented in Table~\ref{tab:text_ns}. For both tasks, omitting the text input resulted in a noticeable increase in WER, whereas the degradation in DNSMOS and PESQ scores was modest. These findings suggest that leveraging the text input was particularly beneficial for enhancing the intelligibility of the output speech. In target speaker extraction, a significant impact on the DNSMOS score was observed, indicating that the text input aids in disentangling the target speaker's voice from the interfering talker. Notably, while relying solely on the acoustic prompt led to WER degradation, the achieved WERs were still comparable to those of the baseline expert models.

\subsubsection{Effect of multi-task training}
Table~\ref{tab:multi-task} shows the results of experiments for assessing the impact of multi-task training. In the first two rows, we present the results of two models
trained from the random initialization. In addition, we trained SpeechX models with single-task training data for clean and noisy speech editing, noise suppression, speech removal, and target speaker extraction tasks, respectively. 
These single-task models served as additional expert models to assess the impact of multi-task training.
The newly trained models were all initialized by the VALL-E checkpoint,
which was trained with the zero-shot TTS task for 400k iterations.
It was then further updated for 400k iterations with single-task training data.

By comparing the first two rows, we observe that the multi-task training significantly improved the WER in the zero-shot TTS task, alongside acquiring the ability to handle additional tasks. This underscores the efficacy of multi-task training, where exposing the model to diverse data can improve single-task performance.
From the lower part of the table, we first observed that 
single-task models trained for noise suppression and target speaker extraction tasks exhibited significantly worse WERs compared to the multi-task model (last row). On the other hand, single-task models trained for speech editing tasks
 showed slightly better WERs than the multi-task model. Other quality metrics such as SIM, DNSMOS, MCD were almost the same.
These observations suggest that the text-conditioned speech generation capability learned through the TTS and speech editing tasks was especially beneficial for achieving high intelligibility in speech enhancement tasks.

We further conducted experiments where we used subsets of the tasks during training to explore potential interactions between different tasks. The results of these experiments are presented in Table~\ref{tab:multi-task-effect}. Specifically, in addition to the VALL-E model and the fully-trained SpeechX models that used the complete set of the tasks, we trained two additional SpeechX models: one trained exclusively for zero-shot TTS and speech editing tasks, and the other trained on the zero-shot TTS, speech editing, noise suppression, and speech removal data. The newly trained models were initialized by VALL-E checkpoint with 400k iterations, and then updated by the subset of the training set. In this experiment, the number of iterations was proportionally reduced based on the number of tasks, with the maximum iteration of 400k with the full set of tasks.

From the results, the inclusion of speech editing during training led to an enhancement in WER for zero-shot TTS while allowing the model to learn about the speech editing task. 
Considering the strong parallels between zero-shot TTS and speech editing, this improvement can be attributed to the speech editing training task introducing additional variations to the distribution of the training data. Further inclusion of the noise suppression and speech removal tasks during training resulted in degradation in clean speech editing performance, while concurrently enhancing the performance for noisy speech editing. This suggests that exposing the model to noisy speech samples from these additional tasks improved the model's robustness to acoustic noise at the expense of clean speech generation. Also, it is noteworthy that introduction of the target speaker extraction tasks to the training data did not compromise the model's proficiency in noise suppression and speech removal. 

\subsubsection{Phoneme vs. Byte Pair Encoding (BPE)}
Table ~\ref{tab:bpe} presents a performance comparison between two variations of the SpeechX model based on the type of textual prompt used: phoneme and BPE. While BPE-based SpeechX performed marginally better in noise suppression and target speaker extraction tasks, phoneme-based SpeechX demonstrated significantly better performance, especially in WER, in zero-shot TTS and clean speech editing tasks. 
For noisy speech editing and speech removal, the phoneme and BPE resulted in similar levels of performance.

We speculate that the severe WER degradation observed with BPE in zero-shot TTS and clean speech editing tasks was due to our utilization of erroneous transcriptions based on ASR during model training. Since BPE units are longer than phonemes, the ASR error rate tends to be higher for BPE. Consequently, when the model was employed for text-based speech generation tasks, where it needed to generate speech from a text prompt, the WER increased significantly.
On the other hand, the text prompt was used only as supplementary information for speech enhancement tasks, such as noise suppression and target speaker extraction. As a result, the WER was not significantly impacted in those cases. It is possible that BPE units may yield better overall performance when the model is trained using ground-truth labels, and this remains an area for future investigation.

\subsubsection{Limitation of current neural codec model}
\label{sec:codec}

\begin{table}[t]
  \centering
  \caption{Impact of neural codec on performance metrics for clean and noisy speech.}
  \tabcolsep = 1.5mm
  \label{tab:codec}
{\footnotesize
  \begin{tabular}{@{}cccccccc@{}}
  \toprule
  \multirow{1}{*}{Audio type}   && { WER$\downarrow$}    & { DNSMOS$\uparrow$} & { PESQ$\uparrow$} & { SIM$\uparrow$}  \\
  \midrule
  \multicolumn{1}{l}{Raw clean speech}   &&  1.71 & 3.22 & 4.64 & 1.00\\
  $\;\;\;\hookrightarrow$ EnCodec  && 1.81 & 2.97 & 2.69 &  0.81 \\ 

  \midrule
  \multicolumn{1}{l}{Raw noisy speech}   && 3.29 &  2.42 & 1.93 & 0.95 \\
  $\;\;\;\hookrightarrow$ EnCodec  && 5.08 & 2.19 & 1.63  & 0.75  \\ 
  \bottomrule 
\end{tabular}
}
\end{table}

The performance of SpeechX is inherently constrained by the accuracy of the neural codec model employed for acoustic tokenization. It should be noted that, in all previous experiments, we compared SpeechX's results with the reference (i.e., no-processing and expert model) results obtained without any neural codec processing. To gain a more precise interpretation of SpeechX's results, we conducted an additional experiment where we applied compression and decompression to the LibriSpeech test-clean data without any intermediate processing, measuring EnCodec's impact on performance metrics.

Table~\ref{tab:codec} shows the experimental results. It is evident that processing the signals with the codec model resulted in varying degrees of performance regression across all metrics. Notably, the PESQ score dropped from 4.64 to 2.69 for the clean speech input. Our assessment indicates that while EnCodec produced slightly noticeable speech quality degradation, the significant PESQ degradation may be partly attributed to the mismatch between the PESQ algorithm and EnCodec's training objective. While we utilized  EnCodec due to its accessibility and prior usage, future work should address this issue by developing an acoustic tokenization model more suitable for handling speech under various acoustic conditions.

\section{Conclusion}

In this paper, we described SpeechX, a novel versatile speech generation model capable of handling diverse audio-text-based speech generation tasks, including zero-shot TTS, noise suppression, speech removal, target speaker extraction, and speech editing. For noise suppression and target speaker extraction, the proposed model provides a unified way for incorporating the knowledge of transcriptions. Also, regarding speech editing, SpeechX enables modifying the spoken content of a speech signal that contains a fair amount of background noise. SpeechX adopts a language modeling approach to generate acoustic tokens conditioned on textual and acoustic prompts, where additional task-dependent tokens are incorporated in a multi-task learning framework to support various speech transformation capabilities beyond zero-shot TTS. We demonstrated SpeechX's efficacy through comprehensive experiments. The proposed model represents an important step toward unified generative speech models. Further research can build on this work by expanding the tasks supported, enhancing robustness and quality (e.g., stabilizing the AR generation with duration control~\cite{song2024ella} and leveraging improved neural codecs~\cite{siuzdak2023vocos}), and developing more advanced conditioning mechanisms (e.g., using disentangled prosody and speaker information~\cite{ju2024naturalspeech}). 

\bibliographystyle{IEEEtran}
{\small\bibliography{refs}}

\vfill

\end{document}